\documentclass[12pt]{iopart}

\usepackage{cite}

\usepackage{graphicx}

\newcommand{\bear}{\begin{eqnarray}}
\newcommand{\eear}{\end{eqnarray}}
\newcommand{\be}{\begin{equation}}
\newcommand{\ee}{\end{equation}}
\newcommand{\beqn}{\begin{eqnarray}}
\newcommand{\eeqn}{\end{eqnarray}}
\newcommand{\beqnn}{\begin{eqnarray*}}
\newcommand{\eeqnn}{\end{eqnarray*}}

\def\vep{\varepsilon}

\begin{document}

\title[Photon statistics in the DCE
modified by a harmonic oscillator detector]
{Photon statistics in the dynamical Casimir effect
modified by a harmonic oscillator detector }

\author{A V Dodonov and V V Dodonov}


\address{
 Instituto de F\'{\i}sica, Universidade de Bras\'{\i}lia,
Caixa Postal 04455, 70910-900 Bras\'{\i}lia, DF, Brazil }

\eads{\mailto{adodonov@fis.unb.br}, \mailto{vdodonov@fis.unb.br}}

\begin{abstract}

It was predicted some time ago that the cavity dynamical
Casimir effect (generation of photons from the initial
vacuum state in a cavity with moving walls)
might be observed if a boundary vibrates
at the double frequency of some selected cavity mode.
However, to register the created photons one has to couple the
cavity mode with some detector. Considering the harmonic oscillator model of
a detector, we analyze how different
coupling regimes can affect the statistics of the created quanta.

\end{abstract}


\section{Introduction}

A possibility of creating quanta of the electromagnetic field from the
initial vacuum state in cavities with moving boundaries, nowadays called 
the Dynamical Casimir Effect (DCE), was a subject of numerous theoretical 
studies for a long time: see, e.g., the most recent reviews
\cite{revDCE,revDal,revRMP}. It was shown \cite{DK96,Plunien,Croc1} that one might expect a considerable
rate of photons generation inside ideal cavities with resonantly oscillating
boundaries.
The simplest model describing this effect takes into account 
 a single resonant cavity mode whose frequency is
rapidly modulated according to the harmonical law $\omega _{t}=\omega
_{0}[1+\varepsilon \sin (\eta t)]$ with a small modulation depth, $%
|\varepsilon |\ll 1$. We shall use
dimensionless variables, setting $\hbar =\omega _{0}=1$. Then the
Hamiltonian for the resonance mode has the form
\cite{Law94}
\begin{equation}
H_{c}=\omega _{t}n-i\chi _{t}(a^{2}-a^{\dagger 2}),
\quad
\chi _{t}=(4\omega _{t})^{-1}d\omega_{t}/dt,
  \label{Hc}
\end{equation}%
where $a$ and $a^{\dagger }$ are the cavity annihilation and creation
operators, and $n\equiv a^{\dagger }a$ is the photon number operator. 
 It is well known that the number of photons created from the
initial vacuum state is maximal if the modulation frequency is 
exactly twice the unperturbed mode frequency, i.e., $\eta =2$. 
The mean number of photons $\langle n\rangle$ and 
the Mandel factor $Q=[\langle (\Delta n)^{2}\rangle -\langle n\rangle
]/\langle n\rangle $  increase with time in this ideal case as
(hereafter we use the subscript $0$ for the quantities related to the empty cavity)
\begin{equation}
\langle n_{0}(t)\rangle =\sinh ^{2}(\varepsilon t/2),
\quad Q_{0}(t)= 
1+2\langle n_{0}(t)\rangle .
  \label{n0}
\end{equation}%
The field mode goes to the squeezed vacuum state
with the following variances of the field quadrature operators 
$x=(a+a^{\dagger })/\sqrt{2}$ and $p=(a-a^{\dagger })/(\sqrt{2}i)$
(in the system rotating with the frequency $\omega_0=1$): 
\begin{equation}
\sigma_{pp}=\frac{1}{2}e^{-\varepsilon t}, \qquad 
\sigma_{xx}=\frac{1}{2}e^{\varepsilon t}.  
\label{p0}
\end{equation}%

But simple formulae (\ref{n0}) and (\ref{p0}) hold for the ideal empty cavity only.
To register the emerging photons one has to couple the field mode to
some detector. And here the problem of the back action of the detector on the field
arises, because in many realistic cases the coupling between the field and detector
can be much stronger than that between the field and vibrating cavity walls.
This was noted  in \cite{pla}, where it was shown that for the
simplest model of detector as a two-level `atom', no photons can be created at all
for the modulation frequency $\eta=2$ if the field--atom coupling constant $g$ is much
bigger than the frequency modulation amplitude $\varepsilon$.
But the photons can be created if one adjusts the modulation frequency 
$\eta = 2(1+ \kappa)$, choosing some nonzero (small) value of
parameter $\kappa$.

Here we consider the model of the detector as a {\em harmonic oscillator\/} 
tuned to the same frequency as the selected field mode. 
Despite its simplicity, this model seems to be rather realistic in the case of the
so called Motion Induced Radiation (MIR) experiment \cite{MIR,MIR2}, where the microwave quanta created via the
DCE are supposed to be detected by means of a small antenna put inside the cavity. 
Since the inductive antenna (a wire loop) used in that experiment is a part of an LC-contour, 
it can be reasonably approximated as a harmonic oscillator. 
Therefore, the Hamiltonian describing the system under study 
(the field mode coupled to such an antenna) can be taken in the form
\be
H= a^{\dagger}a +b^{\dagger}b +g\left(a b^{\dagger } +b a^{\dagger } \right)
-i\chi_t\left(a^2 -a^{\dagger 2}\right)
\label{inHam}
\ee
where the coupling constant $g$ is assumed to be real number. 
\footnote{The term $\omega_{t}n$ in (\ref{Hc}) can be replaced by $n$ because 
for $|\vep| \ll 1$ the main effect of modulation is due to operators $a^{2}$ and $%
a^{\dagger 2}$ in the squeezing part of $H_c$. 
}
Of course, the quadratic Hamiltonian (\ref{inHam}) is an approximation, since it does not
take into account possible nonlinear phenomena, e.g., effects of saturation in the limit of very long
times. Therefore it can be used under the condition $\vep^2 t \ll 1$. But in the present state-of-art experiments 
on DCE, the time scale $\vep t \sim 1$ (or slightly bigger) seems to be quite sufficient for our purposes.

Hamiltonian (\ref{inHam}) contains three real (small) parameters: $g$, $\varepsilon$ and $\kappa$.
Our goal is to find the domains in the space of these parameters where the photon generation
is possible and to study different regimes of generation.
Due to the interaction with the detector, the field mode appears in a mixed quantum state described
by the statistical operator $\hat\rho$. We are interested, in this paper, in the photon distribution
function (PDF) $f(m)\equiv \langle m|\hat\rho| m\rangle$, where $| m\rangle$ means the $m$th Fock state
of the field mode. For the initial vacuum states of the field mode and the detector, the time-dependent 
statistical operator is Gaussian. The general form of PDF 
of the Gaussian states is well known \cite{AgAd,Chat,Mar,MarMar,1mod,book}.
For zero mean values of quadrature components $x$ and $p$,
it can be expressed in terms of the Legendre polynomials as follows \cite{Dod-Turku}:
\be
f(m)=
\frac{2D_{-}^{m/2}}{D_{+}^{(m+1)/2}}
P_m\left(\frac{4\Delta -1}
{\sqrt{D_{+}D_{-}}}\right),
\label{dist0}
\ee
where
\be
D_{\pm} = 1+4\Delta \pm 2\tau,
\qquad \tau = \sigma_{xx}+ \sigma_{pp} \equiv 1 +2\langle n\rangle,
\label{Dpm}
\ee
\be
\Delta = \sigma_{xx}\sigma_{pp} -\sigma_{px}^2 
= \left(\frac12 + \langle a^{\dagger} a\rangle\right)^2 - \left|\langle a^{2}\rangle\right|^2.
\label{def-kap}
\ee 
The Mandel parameter in the Gaussian states with zero first-order moments can be
also expressed through the quantities $\Delta$ and $\langle n\rangle$ as
\be
Q = 1 +2\langle n\rangle - \left(\Delta -1/4\right)/{\langle n\rangle}\,.
\label{Q-Gauss}
\ee
Another quantity we are interested in is the invariant squeezing coefficient 
\cite{book,Dod-Turku,LuPeHr,Loud89,GSQ}
\be
S = \frac{4\Delta}{\tau + \sqrt{\tau^2 -4\Delta}},
\label{S}
\ee 
which does not depend of possible rotations in the quadrature plane, being equal to unity for the
vacuum or coherent states.

\section{Photon generation regimes}

\begin{figure}[htb]
\includegraphics[width=77mm]{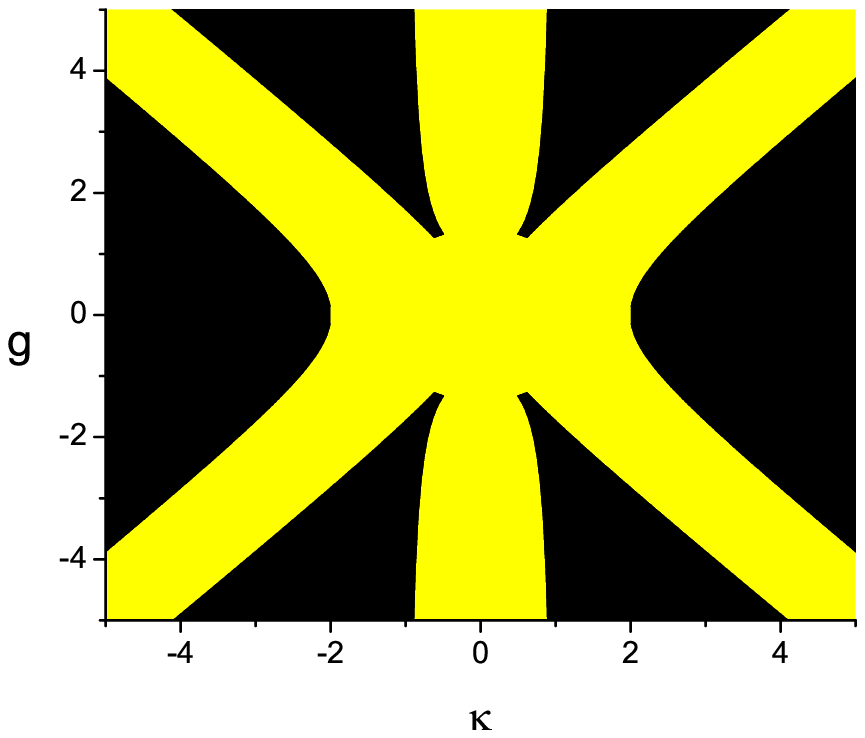}
\includegraphics[width=77mm]{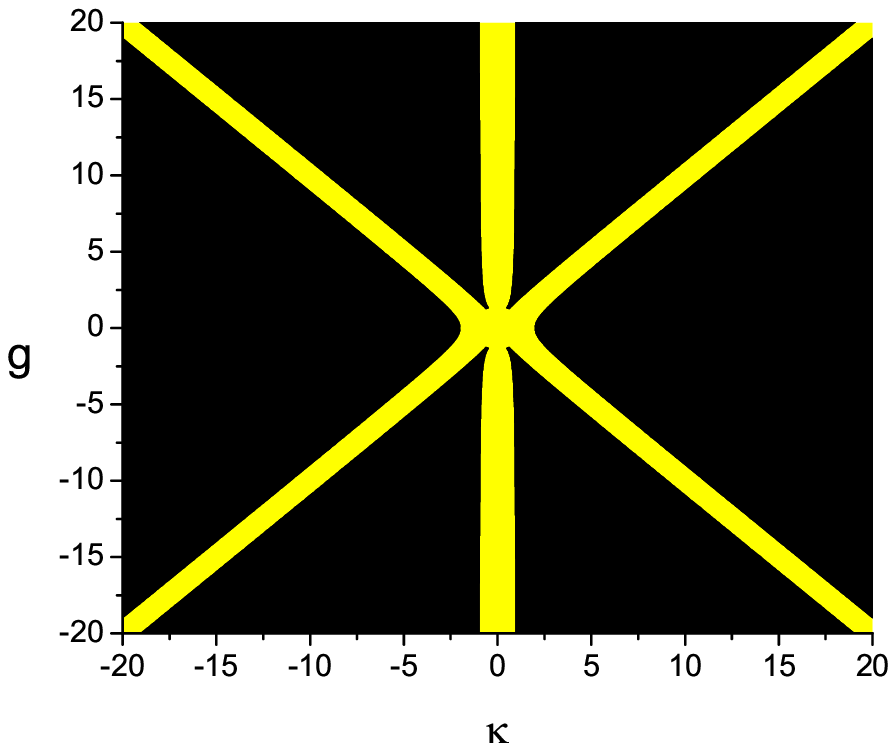}
\caption{The regions (yellow in the online version) in the parameter plane $\kappa$-$g$
where the photon generation from vacuum is possible.
}
\label{Fig1}
\end{figure}

The first two terms in Hamiltonian (\ref{inHam}) can be removed by going to the interaction picture.
Besides, using the rotating wave approximation (RWA) we can remove rapidly oscillating terms
in the product $\chi_t\left(a^2 -a^{\dagger 2}\right)$. Thus, we
arrive at the new Hamiltonian
\be
H_{int}^{(RWA)} = -i\beta\left(a^{2} e^{-2i\kappa t}-a^{\dagger 2} e^{2i\kappa t}\right) + g\left(a b^{\dagger } +b a^{\dagger } \right)
\ee
with $\beta \equiv \varepsilon/4$. The corresponding Heisenberg equations of motion
\be
da/dt = 2\beta a^{\dagger}e^{2i\kappa t} -igb, \qquad
db/dt = -iga
\label{eq-ab-dot}
\ee
can be solved analytically by means of the substitutions
\[
a(t) = e^{i\kappa t} \tilde{a}(t), \qquad
b(t) = e^{i\kappa t} \tilde{b}(t),
\]
which result in equations with constant coefficients
\be
d\tilde{a}/dt = 2\beta \tilde{a}^{\dagger} -igb -i\kappa\tilde{a},
\qquad
d\tilde{b}/dt = -ig\tilde{a} -i\kappa\tilde{b}.
\label{eq-const}
\ee
Looking for solutions to equations (\ref{eq-const}) and their Hermitian conjugated
partners in the form
$\tilde{a}, \tilde{b}, \tilde{a}^{\dagger}, \tilde{b}^{\dagger} \,\sim\, e^{\lambda t}$,
we arrive at the characteristic equation 
\be
\lambda^4 +2\lambda^2\left(\kappa^2 +g^2 -2\beta^2\right) + \left(\kappa^2 -g^2 \right)^2 -4\kappa^2\beta^2 =0
\ee
whose solution reads
\be
\lambda = \pm\sqrt{2\beta^2 -\kappa^2 -g^2 \pm 2\sqrt{\beta^4 -\beta^2 g^2 +g^2\kappa^2}}\,.
\label{lambda-sol}
\ee
The photon generation is impossible if $\mbox{Re}(\lambda_{1,2,3,4})=0$ 
for all four solutions (\ref{lambda-sol}). Otherwise, the real part of at least one characteristic
value $\lambda$ is positive, meaning an exponential growth of solutions.
Analyzing formula (\ref{lambda-sol}) we conclude that the photon generation is impossible
if the following three inequalities are satisfied simultaneously:
\be
\kappa^2 +g^2 > 2\beta^2
\label{cond1}
\ee
\be
\beta^4 -\beta^2 g^2 +g^2\kappa^2 >0
\label{cond2}
\ee
\be
\left(\kappa^2 -g^2 \right)^2 >4\kappa^2\beta^2\,.
\label{cond3}
\ee
If any of the inequalities (\ref{cond1})-(\ref{cond3}) is not satisfied, then an exponential growth
of the mean number of photons can be observed.
In figure \ref{Fig1} we show the regions in the parameter plane $\kappa$-$g$
where the photon generation from vacuum is possible. In this figure all parameters are
normalized by $\beta$ (i.e. formally we put $\beta=1$).

\section{Analysis of special cases}

\subsection{Expected resonances for $|\kappa|=|g|$}
Condition (\ref{cond3}) is obviously broken if $|\kappa|=|g|$, i.e. along the 
bisectrices in figure \ref{Fig1}. The possibility of photon generation in this case 
seems quite natural, as soon as the
corresponding modulation frequency $\eta =2(1+\kappa)$ is exactly twice bigger than one of two
eigenfrequencies  $\omega_{\pm}=1 \pm g$ of the stationary part of Hamiltonian (\ref{inHam})
(with $\chi_t \equiv 0$). Namely, this case was studied for the first time in \cite{DK96}.
In particular, for $|\beta| \ll |g|$ photons can be created if $|\kappa -g|<|\beta|$.
Under this condition the solutions to equations (\ref{eq-ab-dot}) have rather simple
explicit forms if, in addition, $(\beta t)(\beta/g) \ll 1$:
\be
\fl
a(t) = \frac12\left[\left(a_0-b_0\right)\cosh(\beta t) +
\left(a_0^{\dagger}-b_0^{\dagger}\right)\sinh(\beta t)\right]e^{igt} 
+
\frac12\left(a_0 +b_0\right)e^{-igt} ,
\ee
\be
\fl
b(t) = \frac12\left[\left(b_0-a_0\right)\cosh(\beta t) +
\left(b_0^{\dagger}-a_0^{\dagger}\right)\sinh(\beta t)\right]e^{igt}
+
\frac12\left(a_0 +b_0\right)e^{-igt} \,.
\ee
The mean numbers of quanta in both the modes coincide (for the initial vacuum states):
\be
\langle n_a(t)\rangle = \langle n_b(t)\rangle =\frac12\sinh^2(\varepsilon t/4).
\ee
The photon generation rate in the field mode interacting with the oscillator detector 
turns out to be twice smaller than for the empty cavity [given by equation (\ref{n0})].
This result was obtained in \cite{DK96}, but unfortunately the argument of the hyperbolic
sine function there was twice bigger due to a misprint. The effect of diminishing the
photon generation rate due to the resonance intermode interactions was discovered in
\cite{Croc1,DD01}. In the most strong form this effect manifests itself in effectively
one-dimensional Fabry--P\'erot cavities with (quasi)equidistant spectra of eigenfrequencies
\cite{DK96,DD-PLA2012}.
For other statistical properties of the field mode, we have the following formulae:
\be
\Delta =\frac14\cosh^2(\beta t),
\qquad
Q = \frac12\cosh(2\beta t) = 2\langle n\rangle +\frac12,
\ee
\be
\sigma_{xx} = \frac12\cosh(\beta t)\left[\cosh(\beta t) +\sinh(\beta t)\cos(2gt)\right],
\ee
\be
\sigma_{pp} = \frac12\cosh(\beta t)\left[\cosh(\beta t) -\sinh(\beta t)\cos(2gt)\right],
\ee
\be
\sigma_{xp} = \frac12\cosh(\beta t)\sinh(\beta t)\sin(2gt).
\ee
The minimal value (with respect to fast oscillations with frequency $2g \gg \beta$) 
of the variance of any of two quadrature components is equal to
\be
\sigma_{min} = \frac14\left(1 +e^{-2\beta t}\right) \to \frac14\,.
\ee
Formula (\ref{dist0}) for the PDF in the field mode can be
written in the case involved as (see also \cite{DK96})
\be
f(m) = (iz)^m \sqrt{1-3z^2}P_m(-iz), \qquad z\equiv \frac{\tanh(\beta t)}{\sqrt{4-\tanh^2(\beta t)}}\,.
\label{fm-iz}
\ee
For big values of index $m$, one can use the asymptotical formula for the Legendre polynomials
\cite{Olver}
\begin{equation}
P_m(\cosh\xi )\approx
\left(\frac {\xi}{\sinh\xi}\right)^{1/2}
I_0\left(\left[m+1/2\right]\xi\right),
\label{as-Olv}
\end{equation}
where $I_0(z)$ is the modified Bessel function. Taking into account known asymptotical formulae for
the Bessel functions of big (complex) arguments and following the scheme described in \cite{Dod-Turku},
one can arrive at the formula 
\be
f(m) \approx \frac{\displaystyle{
\frac{[\tanh(\beta t)]^m}
{[2-\tanh(\beta t)]^{m+\frac12}} + \frac{[-\tanh(\beta t)]^m}
{[2+\tanh(\beta t)]^{m+\frac12}}}}
{\cosh(\beta t)\sqrt{\pi\left(m+\frac12\right)}},
\label{fm-tanh}
\ee
which is valid under the condition $m\gg 1$ for both small and big values of the product $\beta t$.
It is worth comparing formula (\ref{fm-tanh}) with the strongly oscillating distribution
\be
f_0(2k)= \frac{\langle n\rangle^k (2k)!}{(1+\langle n\rangle)^{k+1/2}(2^k k!)^2},
\quad f_0(2k+1)=0
\label{pdf-sqz}
\ee 
in the squeezed vacuum state arising in the absence of interaction with the detector.
The probabilities of observing odd numbers of quanta in the distribution (\ref{fm-tanh} are
close to zero if $\beta t \ll 1$. But this case is not very interesting, since 
$\langle n(t)\rangle \ll 1$ under this condition. In contrast, if $\beta t > 1$, 
so that $\tanh(\beta t)$ is close to unity and $\langle n(t)\rangle \approx \exp(2\beta t)/8$, 
then one can rewrite (\ref{fm-tanh}) as
\be
f(m) \approx \displaystyle{\frac{
1 + (-1)^m 3^{-m-1/2} }
{\sqrt{2\pi m\langle n(t)\rangle }}}\,.
\label{3-m}
\ee
For $m\gg 1$ the second term in the numerator of fraction in formula (\ref{3-m}) is very small.
Therefore this formula shows very smooth distribution, quite different from (\ref{pdf-sqz}).
Note that for $k \gg 1$ formula (\ref{pdf-sqz}) can be written (using the Stirling formula
for the factorials) as 
$f_0(2k) \approx [\pi k \langle n\rangle]^{-1/2}$. Comparing this expression with (\ref{3-m}) for
$m=2k$ we see that $f(2k) \approx f_0(2k)/2$, so the distribution (\ref{3-m}) can be considered as 
an average of even and odd
values of the `saw-tooth' distribution (\ref{pdf-sqz}).
The plots of exact distributions (\ref{fm-iz}) and (\ref{pdf-sqz}) for $m \le 20$
and  $\langle n\rangle \sim 6$, illustrating these observations,  were given in \cite{DK96}.

On the basis of this example, one could suppose that the drastic change of the behavior of the
PDF is due to the strong coupling with a detector, which plays a role
of some `reservoir' (note that thermal reservoirs usually cause `smoothing' of any
oscillatory behavior). However, the examples of the following subsections show that the real
situation is more intricate, and even the strong coupling with a detector not always destroys
the oscillations of the PDF or some other physical quantities. 
A rough analogy can be the case of nonthermal `rigged' reservoirs,
which can enhance oscillations of some functions.

\subsection{Surprising resonance at $\kappa=0$}
\begin{figure}[tbh]
\includegraphics[width=155mm]{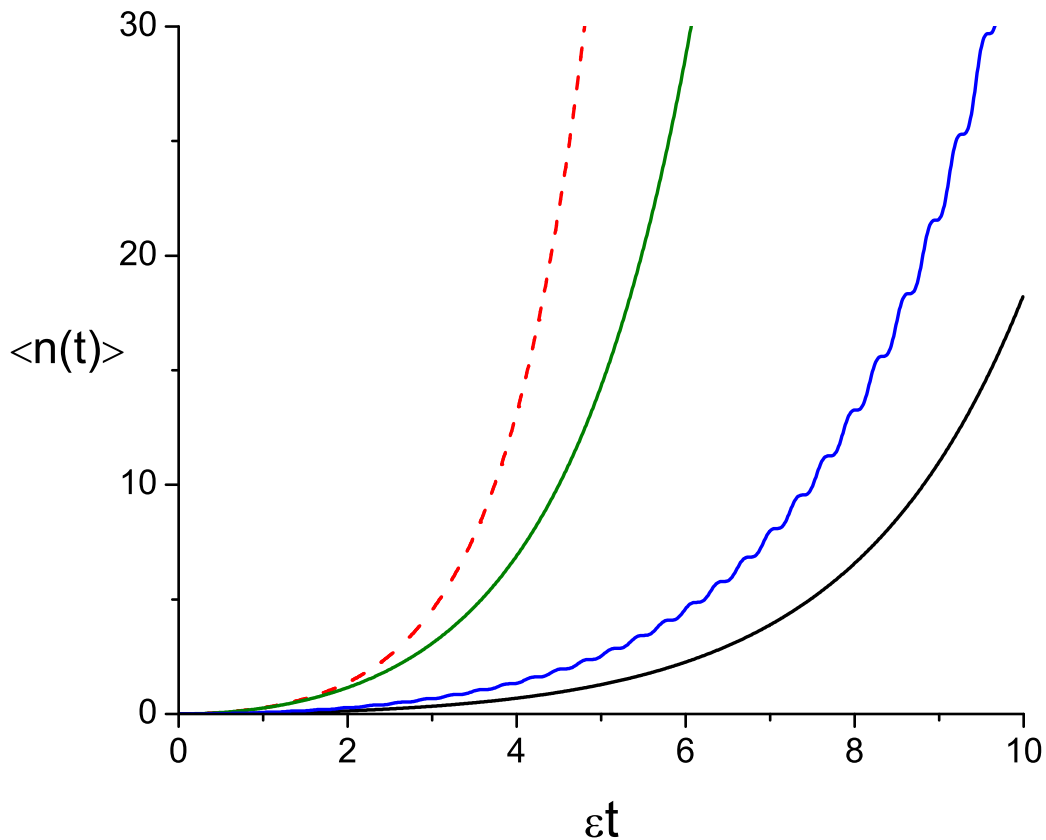}
\caption{The mean number of quanta in the resonance field mode versus the dimensionless
time $\vep t$ for different parameters $\vep$, $g$ and $\kappa$. 
The order of lines from the left to the right: 
$g=\kappa=0$, $\varepsilon=10^{-3}$; $\beta=\kappa=g =10^{-2}$ (see subsection \ref{sub-all});
$\kappa=0$, $g =10 \vep=10^{-2}$; 
$\kappa=g=10^{-2}= 10\vep$.
}
\label{Fig-n}
\end{figure}

Figure \ref{Fig1} shows the existence of resonance photon generation for $\kappa=0$ and for any
value of the coupling constant $g$. This result, first discovered in \cite{DD-PRA-Nlev},
seems surprising, because in the absence of detector (for $g=0$) the mean number of quanta in the case of
a small detuning $\kappa \neq 0$ is given by the following generalization of formula (\ref{n0}):
\be
\langle n_0(t)\rangle = \frac{\vep^2/4 }{\vep^2/4 -\kappa^2}\sinh^2\left(t\sqrt{\vep^2/4 -\kappa^2}\right).
\label{n0kap}
\ee
Formula (\ref{n0kap}) shows that the deviation of the modulation frequency from the resonance value $\eta=2$ by $2\kappa=\vep$
stops the photon generation in the empty cavity. 
Therefore it was natural to expect \cite{DK96} that for $g \neq 0$, the modulation frequency
must be close to $2\omega_{\pm}=2(1\pm g)$, with the deviation not exceeding  something of the order of $\vep$.
Nonetheless, in reality the photons can be created also for $|\kappa| < |\beta|$ even if $|g| \gg |\beta|$.
Perhaps,  this happens due to some kind of quantum interference.
The solutions of equations of motion (\ref{eq-ab-dot}) with $\kappa=0$ and arbitrary values of $\beta$ and $g$ 
can be found in \cite{DD-PRA-Nlev} (similar equations were solved in the contexts of different other physical
problems in \cite{Sete,Zhang}). We bring here only some consequences of that solutions. The mean number of
quanta in the field mode in the case of $|g| \gg |\beta|$ is equal to (for $\beta t \gg 1$)
\begin{equation}
\langle n(t)\rangle  \approx \frac14 e^{2\beta t}\left[ 1
+ \frac{\beta}{\gamma} \sin(2\gamma t) + \frac{2\beta^2}{\gamma^2}\sin^2(\gamma t)\right],
\label{steps}
\end{equation}
where $\gamma =\sqrt{g^2 -\beta^2}$. Again, the rate of photon generation is roughly twice smaller
than in the empty cavity, but the mean photon number is approximately twice bigger than in the case
of $\kappa=g$ considered in the preceding subsection.
Time dependences of the mean numbers of quanta in the field mode in different regimes are
compared in figure \ref{Fig-n}.
The third (blue) line from the left (corresponding to the case of $\kappa=0$) shows remarkable
horizontal steps. This peculiar behavior was explained in \cite{DD-PRA-Nlev}.

Formula (\ref{dist0}) indicates that even-odd oscillations of the PDF can happen if
the argument of the Legendre polynomial is close to zero [since $P_{2m+1}(0) = 0$ while 
$P_{2m}(0)=(-1)^m (2m-1)!!/(2m)!!$]. In the most strong form these oscillations manifest themselves for pure
quantum Gaussian (squeezed) states with $4\Delta -1 \equiv 0$, as one can see in formula (\ref{pdf-sqz}).
In the case of $\kappa=0$ we have $4\Delta -1 = 4g^2 \beta^2 \sin^4(\gamma t)/\gamma^4$, and this quantity is 
very small if $|g|\gg |\beta|$. Therefore, contrary to the case of $\kappa=g \gg \beta$, the PDF shows 
oscillations, and for $m\gg 1$ we have
\be
f(2m) = \frac{\tanh^{2m}(\beta t)}{\cosh(\beta t)}\frac{(2m-1)!!}{(2m)!!},
\ee
\be
\frac{f(2m+1)}{f(2m)} =\frac{2(2m+1)\beta^2 \sin^4(\gamma t)}{g^2 \sinh(2\beta t)}
\ll 1.
\ee

\subsection{The intermediate regime $\kappa=g=\beta$}
\label{sub-all}

Another interesting special case admitting explicit solutions is $\beta=\kappa =g$
(the point of intersection of the bisectrice of the first quadrant and the circle
$\kappa^2 +g^2 =2\beta^2$ [see condition (\ref{cond1})] in figure \ref{Fig1}). The characteristic values
are $\lambda_{1,2} = \pm \sqrt{2}\,\beta$ and $ \lambda_{3,4} =\pm i\sqrt{2}\,\beta$, so that
the following exact formulae hold ($x=\sqrt{2}\,\beta t$):
\beqnn
a(t) &=&\frac{e^{i\beta t}}{\sqrt{8}}\Bigg\{a_0\left[\sqrt{8}\,\cosh(x) -i\sinh(x) -i\sin(x)\right]
 \\ &&
+a_0^{\dagger}\left[3\sinh(x) +\sin(x)\right]
 \\ && 
+ b_0 \left[\sqrt{2}\,\cos(x) -\sqrt{2}\,\cosh(x) -i\sinh(x) -i\sin(x)\right]
\\ &&
+ b_0^{\dagger} \left[ \sin(x) -\sinh(x) +i\sqrt{2}\,\cosh(x) -i\sqrt{2}\,\cos(x) \right]
\Bigg\},
\eeqnn
\be
\langle a^{\dagger} a\rangle = \frac12\left[ 1 +3\sinh^2(x) +\sinh(x)\sin(x) -\cosh(x)\cos(x)\right].
\label{a+a}
\ee
Despite the presence of trigonometric functions in formula (\ref{a+a}), the mean number of photons
grows practically exponentially without visible oscillations, as shown by the second line from the left in
figure \ref{Fig-n}. The asymptotical rate of photon generation in this case is equal to $\vep/\sqrt{2}$ --
an intermediate value between $\vep$ and $\vep/2$ characterizing the two adjacent curves.

The parameters entering formula (\ref{dist0}) for the PDF are as follows:
\beqnn
4\Delta -1 &=& 2\cosh^2(x) -2 \left[\sinh(x)\sin(x) +\cosh(x)\cos(x)\right]
\\ &&
-\frac12 \left[\cosh(x)\sin(x) - \sinh(x)\cos(x) \right]^2,
\eeqnn
\[
D_{+} = 8\cosh^2(x) -4 \cosh(x)\cos(x) 
-\frac12 \left[\cosh(x)\sin(x) - \sinh(x)\cos(x) \right]^2,
\]
\[
D_{-} = -4\sinh^2(x) -4 \sinh(x)\sin(x) 
-\frac12 \left[\cosh(x)\sin(x) - \sinh(x)\cos(x) \right]^2.
\]
For $x\gg 1$ we have
\[
4\Delta -1 \approx \frac12e^{2x}(1-\xi/4), 
\qquad \xi =(\sin\,x -\cos\,x)^2, 
\]
\[
D_{+} \approx 2e^{2x}(1-\xi/16), \qquad
D_{-} \approx -e^{2x}(1+\xi/8).
\]
Consequently,
\[
f(m) \approx \sqrt{2}\left(\frac{1+3\xi/16}{2}\right)^{m/2} e^{-x}\, i^m P_m\left(-\frac{i(1-\xi/4)}{\sqrt{8+\xi/2}}\right).
\]
Since $ 0\le \xi \le 2$, the argument of the Legendre polynomial varies from $-i/\sqrt{8}$ to $-i/6$, i.e. it cannot assume 
very small values. Therefore the PDF does not show noticeable oscillations, and can be well approximated
(for $1\ll m \sim \langle n\rangle$) by the
formula \cite{Dod-Turku,D-PRA09}
\be
f(m) \approx \frac{\exp\left[-(2m+1)/(4\langle n\rangle)\right] }{\sqrt{\pi\langle n\rangle(2m+1)}}\,.
\ee
The quantity showing oscillations in the case concerned is the invariant squeezing coefficient 
$S=2\sigma_{min}$. Indeed, since $\Delta \sim \tau \gg 1 $ for $x \gg 1$, formula (\ref{S}) can be
simplified as $S\approx 2\Delta/\tau$, so that 
$ S(x\gg 1) \approx (1-\xi/4)/{3}$. Since $\xi(x)$ is a periodic function of time, the minimal quadrature variance
$\sigma_{min}$ does not go asymptotically to some limit, but it oscillates between the values $1/6$ and $1/12$.

\section{Conclusions}
The main results of this paper are as follows. We found the conditions of photon generation
in a three-dimensional cavity with resonantly oscillating ideal walls when the resonance field mode
is linearly coupled to a detector modeled as harmonic oscillator. The `allowed' and `forbidden' zones
in the space of parameters $\vep$, $g$ and $\kappa$ are presented in figure \ref{Fig1}. 
We have shown that the main physical
observables, such as the mean number of created quanta, their distribution function and the invariant
squeezing coefficient, can show either smooth monotonic behavior or some kinds of oscillations,
depending on the parameters  characterizing the process. However, the oscillations
of different quantities seem to be uncorrelated, according to the examples considered.

\section*{Acknowledgments}
AVD and VVD acknowledge the support of the Brazilian agencies CAPES
and CNPq, respectively.

\end{document}